\documentclass[12pt]{iopart}

\usepackage[pdftex]{graphicx} 
\usepackage{epstopdf}
\usepackage{iopams}
\usepackage{color}  
\begin{document}

\title[J. Phys.: Condens. Matter]{Theoretical Investigations of  Electronic Structure, Magnetic and Optical Properties of Transition Metal-dinuclear Molecules}

\author{Indukuru Ramesh Reddy \& Kartick Tarafder*}

\address{Department of Physics, National Institute of Technology Karnataka, Srinivasnagar, Surathkal, Mangalore, Karnataka-575025, India}
\ead{karticktarafder@gmail.com}
\vspace{10pt}
\begin{indented}
\item[]\today
\end{indented}

\begin{abstract}
The work presents the electronic structure, spin state and optical properties  of  TM-dinuclear molecules (TM = Cr, Mn, Fe, Co, and Ni) which was modelled according to the recently reported Pt$^{II}$-dinuclear complex data\cite{kar_colour_2016}. The molecules were geometrically optimized in the gas phase and their stability were analyzed from  vibrational spectra study using density functional theory (DFT) calculations. The ground spin state of the tetra-coordinated TM atom in the modeled molecules was predicted based on the relative energies between the possible spin states of the molecules. We further performed DFT+U calculations to investigate the precise ground state spin configuration of molecules. Interestingly, optical characterization of these molecules shows that  the absorption spectra have a large peak in the blue-light wavelength range, therefore could be suitable for blue-LED application. Our work promotes further computational and experimental studies on TM-dinuclear molecules in field of molecular spintronics and optoelectronics. 
\end{abstract}

\vspace{2pc}
\submitto{\JPCM}

\section{Introduction}
Quinonoids are promising molecules, well known for their application  in physical chemistry, color chemistry and in several biological applications\cite{patai_chemistry_1988,lekin_benzoquinone-bridged_2018}.  This class of molecule and their derivatives are highly pH-selective and pH-sensitive \cite{elhabiri_acidbase_2004,su_19-dihydro-3-phenyl-4h-pyrazolo[34-b]quinolin-4-one_2001,izumi_p-quinone_2005}. Their  presence in organic thin films singnificantly enhances the charge carrier concentrations. As a result highly conductive interface can be formed by adsorbing these molecule and their derivatives on  metal/semiconducting  surfaces \cite{routaboul_altering_2012}. These attribute that considering  quinonoids and their derivatives as  active centers, one can take part in developing new functional materials. In addition, metal-organic quinonoid molecules are potential building blocks in  coordination polymers as well as materials useful for catalysis, surface chemistry, and molecular spintronics\cite{kim_multifunctionality_2013,dei_quinonoid_2004}. In case of  transition metal-quinonoid(TM-quinonoid), the valence electrons, which are primarily from partially filled d-orbitals,  stabilizes in various electronic spin-configurations in different situations. That allow us a reversible spin state switching in the system by using suitable external parameter such as temperature, pressure, electric and magnetic field. Molecule which exhibits reversible spin switching are useful for memory storage devices and molecular switches. Mixed valence and valence tautomerism in TM-quinonoid complexes have already been reported in molecular switching applications\cite{dei_quinonoid_2004}. Recently a multifunctional Ni-quinonoid complex has been reported in which a Ni atom connects two quinonoid molecules, exhibits a reversible color change and undergoes structural transformation upon methanol adsorption\cite{kar_methanol-triggered_2017}. In one of our previous report we have shown that the Ni-quinonoid chemisorbed on Co(100) surface, couple ferromagnetically with substrate, in which Ni spin-state switches from Low Spin (LS) to High Spin (HS) \cite{reddy_interfacial_2019}. Several quinonoid based organometalic compounds have been synthesized in the past few year aiming to their application in spintronics. Kar et. al. have recently reported quinonoid based homo- and hetero-dinuclear complexes where the two noble metals (Pt$^{II}$ and/or Pd$^{II}$) are  linked through a quinonoid molecule. Herein, we modeled  TM homo-dinuclear organometallics by replacing both the noble metals in the molecule with different transition metal atoms and carried out theoretical investigations of their stability, electronic structure,  magnetic and optical properties using first principles density functional theory calculations.

\section{Computational Methadology}
TM-dinuclear molecules were modelled according to the reported structure of the Pt$^{II}$-dinuclear complex\cite{kar_colour_2016}. Electronic structure and optical absorption spectra of modelled TM-dinuclear molecules were obtained by using Density Functional Theory (DFT) calculations implemented in GAUSSIAN 09 package. All-electron Gaussian basis set developed by Ahlrichs and co-workers was employed for all the elements \cite{schafer_fully_1994} in each system. Prediction of proper ground state spin configuration  of  organometallic molecules is a challenging task as the total energy difference is very small for such complex in different spin states configuration\cite{neese_efficient_2009}. Recent studies showed that the meta-GGA hybrid fucnctional (TPSSh functional)  which uses 10\% Hartree-Fock exchange to approximate the  exchange correlation functional for TM atoms, works excellently and predicts accurate spin configurations in  TM-organometallic complexes\cite{jensen_accurate_2009}. Hence, we have used the TPSSh functional with the large triple-$\zeta$ basis set along with  polarization function for all the atoms. The geometric structure of all the molecules were optimized using this functional. Bernys optimization algorithm which involves redundant internal coordinates, was used to optimize the molecular geometry.
The detail magnetic structure of TM-organometallics were obtained using the plane wave, pseudopotential method as implemented in Vienna ab initio simulation package (VASP)\cite{kresse_efficient_1996} with projected augmented plane wave (PAW) potential\cite{blochl_projector_1994}. Generalized Gradient Approximation (GGA) was used with Perdew-Burke-Ernzerhof parameterization for exchange-correlation function\cite{perdew_generalized_1997}. We used DFT+U technique to get the precise spin state of the organometallic by capturing the strong electron-electron correlation effect in the partially filled 3d shell of TM atoms and missing correlation effect beyond GGA\cite{tarafder_pressure_2012,reddy_route_2018}. We have tested with different on-site Coulomb parameter (U) and exchange parameter (J) for TM atoms. A detail description is given in the Supporting Information (SI). A plane wave kinetic energy cutoff 500 eV was considered. The convergence criterion was set to 10$^{-5}$ eV for the self-consistent electronic minimization. Forces on each atom are estimated using  Hellman Feynman theorem and subsequently structural optimization was carried out using conjugate gradient method until force on each atom was reduced to 0.01 eV/\AA.
\section{Results and Discussion}
\subsection{Optimized geometric structure :}
\begin{figure}[h!]
 	\includegraphics[width=16cm]{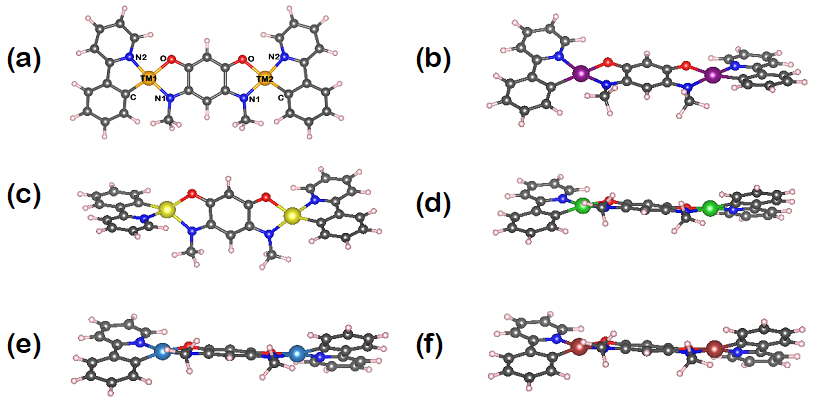}
 	\caption{(color online)(a) Modelled structure of the TM-dinuclear molecule, TM, O, N, C, and H atoms are represented by orange, red, blue, black, and light pink colored balls, respectively. (b-f) Optimized structure of the Cr$^{II}$-, Mn$^{II}$-, Fe$^{II}$-, Co$^{II}$-, and Ni$^{II}$-dinuclear molecules. Cr, Mn, Fe, Co and Ni atoms are represented by magenta, yellow, turquoise and bronze, respectively, whereas other atoms are shown same as in (a).}
 	\label{fig1}
 \end{figure}

 The molecular structure of TM-dinuclear molecule consist of two transition metal (TM) one quinonoid and two bi-pyridine ligands.  Each one of two TM atoms in the molecule is attached with a bi-pyridine ligand and they are connected through a quinonoid. The modeled structure of TM-dinuclear molecule  is shown in Fig.{\ref{fig1}a}. In order to proceed, we first optimized the geometric structure and  calculated the vibrational spectra for each modeled molecule to test the stability. The calculated IR  spectra of TM-dinuclear molecules are given in Fig.SI-1. Absence of negative frequencies in vibration spectra confirms that the optimized structure of molecules are stable.  We observed slight differences in optimized  molecular geometries.  The average TM-ligand bond lengths ($\Delta$TM-L) are 2.05 \AA, 2.09 \AA, 1.95 \AA, 1.93 \AA \ and 1.90\AA \   for Cr, Mn , Fe , Co and Ni dinuclears  respectively.  Distortion of planer structure around TM-atom site  is also different in different  molecule depending on the TM atom.  
 The distortion from a square-planar to a tetrahedral structure can be indexed with $\tau_{4}$ parameter, $\tau_{4}$ is 0 for the perfect square-planar and 1 for the perfect tetrahedral geometry. The $\tau_{4}$ parameter for the distortion in the tetra-coordinated TM of the molecules is tabulated in Table [\ref{t3}]. (See SI for the calculation of $\tau_{4}$ parameter). 
 \begin{table}
 	\caption{\label{t3}The calculated average TM - Ligand bond lengths (in Å) and the distortion parameter in the optimized geometry of TM-dinuclear molecules.}
 	\begin{indented}
 		\item[]\begin{tabular}{@{}llllll}
 			\br
 			Molecule& TM& Average TM - Ligand bind length in (Å)& ~~~~~~~~~$\tau_{4}$\\
 			&   & ~~~~~~TPSSh~~~~~~~~~~~~~~~GGA+U& TPSSh& GGA+U\\
 			\mr
 			$Cr^{II}$-& Cr-1& ~~~~~~2.05847~~~~~~~~~~~~~~~2.07833&	0.16&	0.16\\
 			& Cr-2& ~~~~~~2.05846~~~~~~~~~~~~~~~2.07840&	0.16&	0.16\\
 			$Mn^{II}$-& Mn-1& ~~~~~~2.09841~~~~~~~~~~~~~~~2.14652&	0.75&	0.38\\
 			& Mn-1& ~~~~~~2.09840~~~~~~~~~~~~~~~2.14627&	0.75&	0.38\\
 			$Fe^{II}$-& Fe-1& ~~~~~~1.95588~~~~~~~~~~~~~~~1.96381&	0.15&	0.16\\
 			& Fe-2& ~~~~~~1.95589~~~~~~~~~~~~~~~1.96376&	0.15&	0.16\\
 			$Co^{II}$-& Co-1& ~~~~~~1.93057~~~~~~~~~~~~~~~1.92874&	0.17&	0.17\\
 			& Co-2& ~~~~~~1.93058~~~~~~~~~~~~~~~1.92900&	0.17&	0.17\\
 			$Ni^{II}$-& Ni-1& ~~~~~~1.90151~~~~~~~~~~~~~~~1.90319&	0.16&	0.16\\
 			& Ni-2& ~~~~~~1.90148~~~~~~~~~~~~~~~1.90323&	0.16&	0.16\\
 			\br
 		\end{tabular}
 	\end{indented}
 \end{table}
 
 It shows that the tetra-coordinated Mn atoms in  Mn$^{II}$-dinuclear molecule is more distorted towards tetrahedral geometry compare to other TM-dinuclear molecules, which are nearly in the square-planar geometry. According to the well-known Irving-Williams order\cite{zhang_electronegativities_1982}, the stability of the TM increases in the order Mn$^{II}<$ Fe$^{II}<$ Co$^{II}<$ Ni$^{II}$. Therefore, the distortion in the Mn$^{II}$-dinuclear molecule could be ascribed due to less ligand field stabilization energy\cite{li_estimation_2006}. Optimized geometry for each of these molecules are shown in Fig[\ref{fig1}b-f].
 
 \subsection{Electronic structure and magnetic properties}
 The magnetic behavior of TM-dinuclear molecules depends on the co-ordination of TM and their local environment.   In general, tetra-coordinated (square-planarar) TM centers with more than half field d-orbitals do not exhibit the HS state due to a large John-Tellar effect[10]. Therefore, depending on the number of electrons present in d-orbitals of transition metal atoms, the spin state of the modeled organometallic molecules are expected to  be either intermediate spin (IS) or  LS. Hence one of  singlet, triplet, and quintuplet would be the possible spin states (or spin-multiplicities) for the molecule with an even number of electrons in the d-orbitals, whereas the doublet, quartet, or sextet would be the  possible spin states for those with an odd number of electrons in the d-orbitals. The ground state spin-configuration of the organometallic molecules was evaluated by calculating the total energy of the molecule in all the possible spin states. The  difference in total energy with respect to the ground state energy in different spin states are listed in Table-\ref{t1}.  Our calculations shows that the Cr$^{II}$- and Mn$^{II}$-dinuclear molecules are in high spin configuration, Fe$^{II}$-dinuclear  is in intermediate spin configuration , Co$^{II}$- and Ni$^{II}$-dinuclear molecules are in low spin state  in their respective ground state. 
 Average TM-ligand bond lengths obtain in our calculation also supports the predicted spin state of  molecules. \cite{tarafder_pressure_2012,hauser_ligand_2004}.
 %
\begin{table}
\caption{\label{t1}The calculated total energy difference in the possible spin states with respect to the ground spin state energy of the TM-dinuclear molecules in eV. Total energy=0 eV represents the ground state spin configuration. Corresponding  spin-multiplicity of TM is shown in  parentheses}
\begin{indented}
\item[]\begin{tabular}{@{}llll}
\br
Molecule& \multicolumn{3}{c}{Relative total energy and Spin state(in eV)}\\
		&~~~~~LS &~~~~~IS &~~~~~ HS\\
\mr
$Cr^{II}$-& 5.08 (singlet)& 1.263 (triplet)& 0 (quintuplet)\\
$Mn^{II}$-& 2.19 (doublet)& 0.296 (quartet)& 0 (sextet)\\
$Fe^{II}$-& 2.22 (singlet)& 0 (triplet)& 0.92 (quintuplet)\\
$Co^{II}$-& 0 (doublet)   & ~~~~~~- & 0.63 (quartet)\\
$Ni^{II}$-& 0 (singlet)   & ~~~~~~- & 1.86 (triplet)\\
\br
\end{tabular}
\end{indented}
\end{table}

In order to have a clear insight into the predicted spin state, we further performed the GGA+U calculations and obtained detail atom specific spin moments of molecules. The obtained magnetic moment on the both transition metal atoms in each molecule are listed in Table [\ref{t2}]. 
\begin{table}
	\caption{\label{t2}Calculated magnetic moments (in $\mu_{B}$) and spin state (in parenthesis) on the TM atoms in TM-dinuclear molecules.}
	\begin{indented}
		\item[]\begin{tabular}{@{}lll}
			\br
			Molecule&\multicolumn{2}{c}{Magnetic Moment}\\
			&TM-1& TM-2\\
			\mr
			$Cr^{II}$-& 3.730 (S=2)  & 3.730 (S=2)\\
			$Mn^{II}$-& 4.598 (S=$\frac{5}{2}$)& 4.598 (S=$\frac{5}{2}$)\\
			$Fe^{II}$-& 2.149 (S=1)  & 2.149 (S=1)\\
			$Co^{II}$-& 1.083 (S=$\frac{1}{2}$)& 1.083 (S=$\frac{1}{2}$)\\
			$Ni^{II}$-& -0.012 (S=0) &-0.012 (S=0)\\
			\br
		\end{tabular}
	\end{indented}
\end{table}

\begin{figure}[h!]
	\includegraphics[width=16cm]{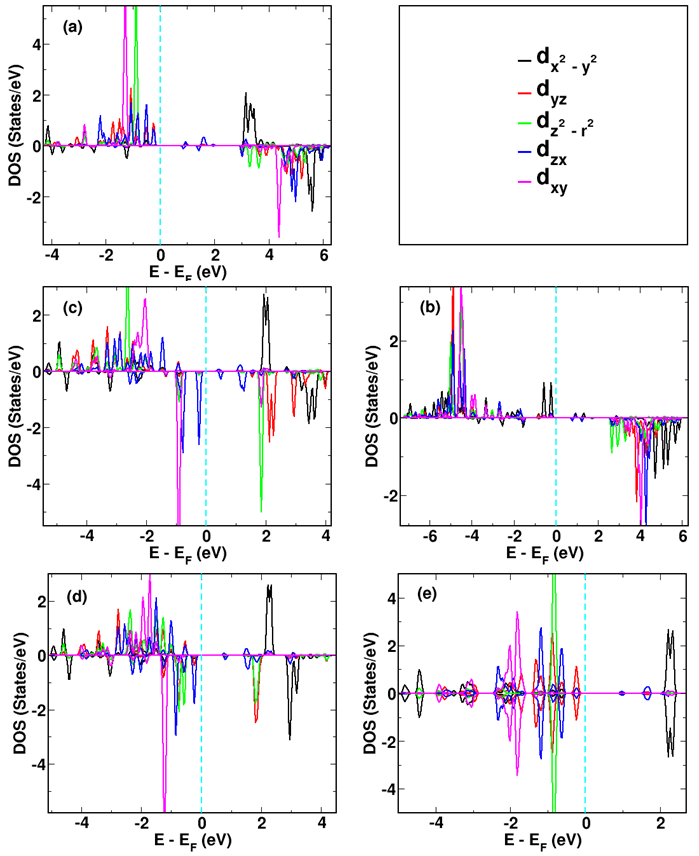}
	\caption{(color online) The d-orbital projected DOS of TM atom in: (a) Cr$^{II}$-, (b) Mn$^{II}$-, (c) Fe$^{II}$-, (d) Co$^{II}$-, and (e) Ni$^{II}$-dinculear molecule. Spin-up and spin-down DOS are shown by positive and negative values respectively.}
	\label{fig2}
\end{figure}
To understand the electrons occupation in the d-orbitals, we have obtained the spin and d-orbital resolved-Density of States (DOSs) for each molecule, plotted in Fig[\ref{fig2}a-e]. TM atoms in molecules are tetra-coordinated and bonded with O, N and C  atoms of molecular ligands in the xy-plane. Therefore $\sigma$-type hybridization between the in-plane 3$_{d}$ and 2$_{p}$ orbitals will originate crystal field effect in d-orbitals of TM atoms.  The coulomb repulsion along with strong $\sigma$-type hybridization will lead to a splitting of d-orbital into t$_{2g}$ and e$_{g}$ orbitals. On the other hand, in absence of vertical co-ordination of TM, the degenerated $eg$ orbitals will further split into two states, in which  d$_{z^2-r^2}$ orbital will be  lower in energy compared to  d$_{x^2-y^2}$. 
 
Relatively shorter TM-ligand  bond lengths in Fe-,Co- and Ni-dinuclear  will induce strong ligand field and affect the in-plane d-orbitals of TM atom that will further rise the in-plane orbital energy levels. As a result, electron distribution in d-orbital for 3d$^{6}$-3d$^{8}$ electronic configurations will  violate the Hund’s rule. Electrons will be forced to occupy the  d$_{xz}$, d$_{yz}$, d$_{z^2-r^2}$ and d$_{xy}$ orbitals first to saturate these orbitals, leaving the d$_{x^2-y^2}$ orbital completely unoccupied. The d-orbital projected density of state for the system with 3d$^{6}$-3d$^{8}$ configurations are shown in the Fig.[\ref{fig2}-a,c,d \& e], which clearly shows that both the up and down spin  d$_{x^2-y^2}$ states are in the conduction band. In the case of Fe$^{II}$-dinuclear complex, the d$_{xz}$ and d$_{xy}$ orbitals are completely occupied, d$_{yz}$ and d $_{z^2-r^2}$ orbitals are partially filled with spin-up electrons. As a result, the spin state of the Fe$^{II}$ atom turned out to be S=1 (See Fig. 2c.). In a similar way, only d$_{yz}$ orbital in Co$^{II}$-complex is  partially filled with spin-up electron and all three d$_{xz}$, d$_{xy}$, d$_{z^2-r^2}$ orbitals are completely occupied. Thus the spin state of the Co$^{II}$ atom is  S=1/2 (See Fig. 2d.). In case of Ni{$^{II}$}-dinuclear all four d$_{xz}$, d$_{yz}$, d$_{z^2-r^2}$ and d$_{xy}$ orbitals are completely occupied and d$_{x^2-y^2}$ orbital is completely unoccupied
 leads to S=0 spin state for this system (See Fig. 2e.), where as in case of Cr$^{II}$-dinuclear these four orbitals are partially occupied with the spin-up electrons gives rise to   S =2 (See Fig. 2a.) spinstate in this system. Interestingly, Mn$^{II}$-dinuclear does not follow this rule. Since the tetra-coordinated Mn atoms in Mn$^{II}$-dinuclear molecule are more distorted towards tetrahedral geometry, also the average  TM-Ligand bond length is   increased. Therefore  realignment of  the d-orbitals according to the tetrahedral co-ordination is expected. As a result, all the d-orbitals including in-plane d$_{x^2-y^2}$ orbital of Mn in the Mn$^{II}$-dinuclear molecule are  singly occupied with spin-up electrons, and the spin-down channel is completely empty. Hence, the spin state of the tetra-coordinated Mn$^{II}$ atom in the molecule is turned out to be S=5/2 (See Fig. 2b.).\\

\subsection{Optical properties}
\begin{figure}[h!]
\includegraphics[width=0.75\linewidth]{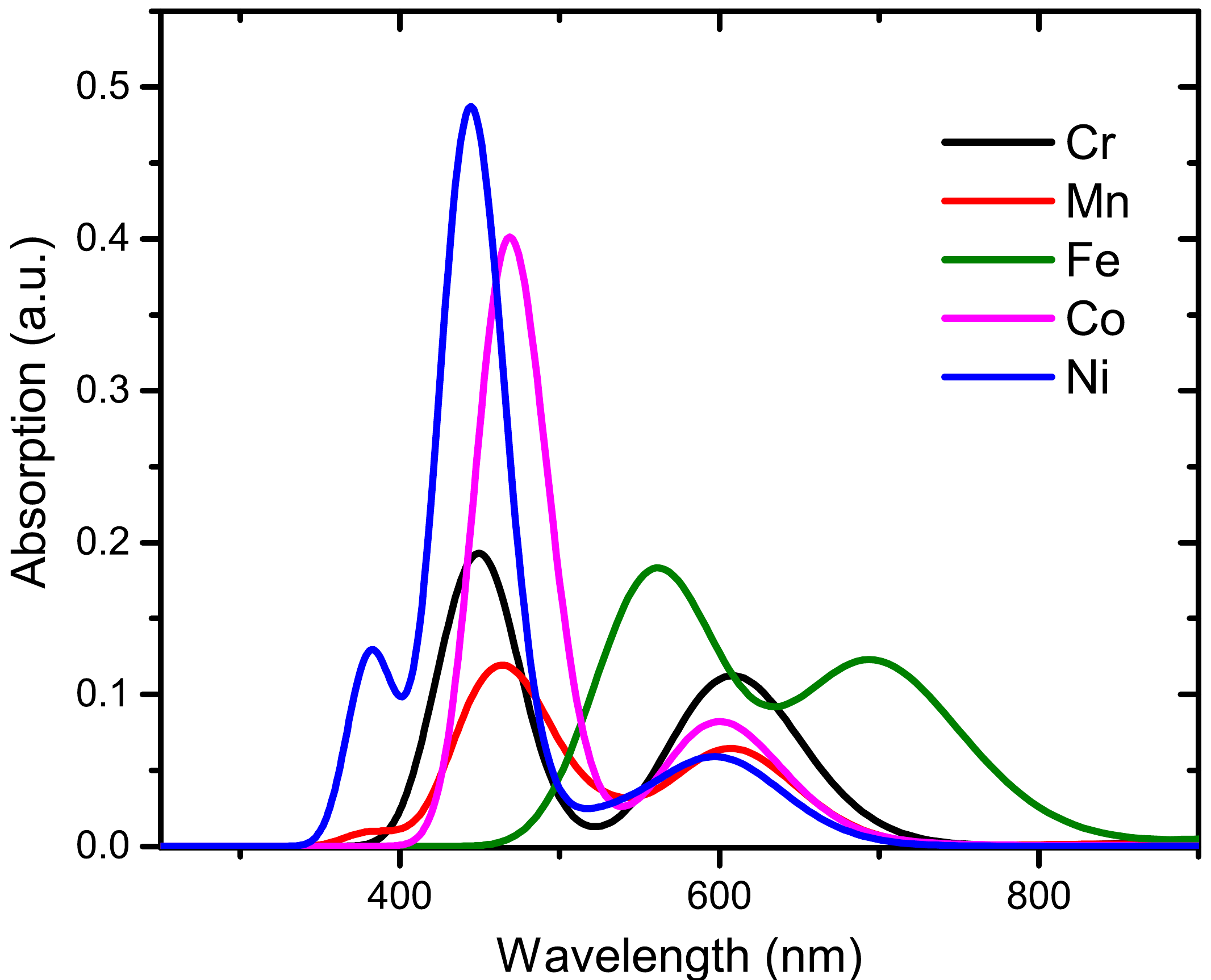}\quad
	\caption{(color online) UV-Vis absorption spectra of all TM-dinuclear molecules obtained from TD-DFT calculations.}
	\label{fig3}
\end{figure}

 Optical response is a useful tool to understand the proper electronic behavior of molecules. Therefore we have performed time-dependent density functional theory (TD-DFT) calculations based on the optimized structure of TM-dinuclear molecules by using B3LYP functional and calculated electronic transitions. The UV-visible optical absorption spectra for each molecule  is plotted in Fig.[\ref{fig3}]. 
We observed that there are two major peaks in the absorption spectra. First one is in the range of 400-500nm which has maximum intensity  and the second one is in the rage of  550-650 nm which is less intense and wide. The position of  the  maximum intensity peaks, band transitions energy and transition probabilities of TM-dinuclear molecules are listed in the Table [\ref{t4}].
\begin{table}
	\caption{\label{t4}Calculated transition energies, absorption wavelength and corresponding transition probabilities for TM-dinuclear molecules.}
	\begin{indented}
		\item[]\begin{tabular}{@{}llll}
			\br
			Molecule& Transition Energy(eV)& Transition wavelength($\lambda$) & Transition Probability\\
			\mr
			$Cr^{II}$- & 2.7029 &	458.71 nm & 0.1267  \\
			$Mn^{II}$- & 2.7801 &	476.55 nm & 0.0889  \\
			$Fe^{II}$- & 2.5287 &   490.31 nm & 0.1488  \\		
			$Co^{II}$- & 2.6453 &   468.69 nm & 0.4268  \\	
			$Ni^{II}$- & 2.7877 &   444.76 nm & 0.4624  \\
			
			\br
		\end{tabular}
	\end{indented}
\end{table}
Details of the other relevant  transitions are given in SI. Maximum intensity of the first peak is observed from Ni-quinonoid molecule. A similar but slightly less intensity peak is found for Co-molecule. For other three TM-dinuclear molecule the absorption band is quite broad. In case of Fe-molecule the first peak appears near 550nm wavelength region. From the optical absorption spectra we can predict that dinuclear molecule containing Cr, Ni, Co and Mn  atoms would be useful for blue light emitting diode applications, in which Ni-dinuclear molecule will show maximum efficiency.  Whereas Fe-dinuclear would be useful for yellow light emission. To understand the nature of inter/intra band transition under the influence of photon absorption, we plotted upto seven upper-most occupied molecular orbitals(HOMO to HOMO-6) and seven lowermost unoccupied molecular orbitals(LUMO to LUMO+6) for each system. Details of molecular orbitals are given in SI.
Molecular orbital plots are clearly show that  $\pi$-electrons in quinonoid (Q$^{2-}$)  adopt a zwitterionic form with two delocalised subunits i.e., trimethine oxonol (OCCCO) and trimethine cyanine (NCCCN) in each molecule. Higher occupied orbitals specially HOMO and HOMO-1 are mainly contributed either from  NCCCN or OCCCO subunit orbitals in most of the cases. On the other hand lower unoccupied orbitals are localized mainly on the electron deficient C–C single bond of the quinonoid ligand.  
\begin{figure}[h!]
	\includegraphics[width=16cm]{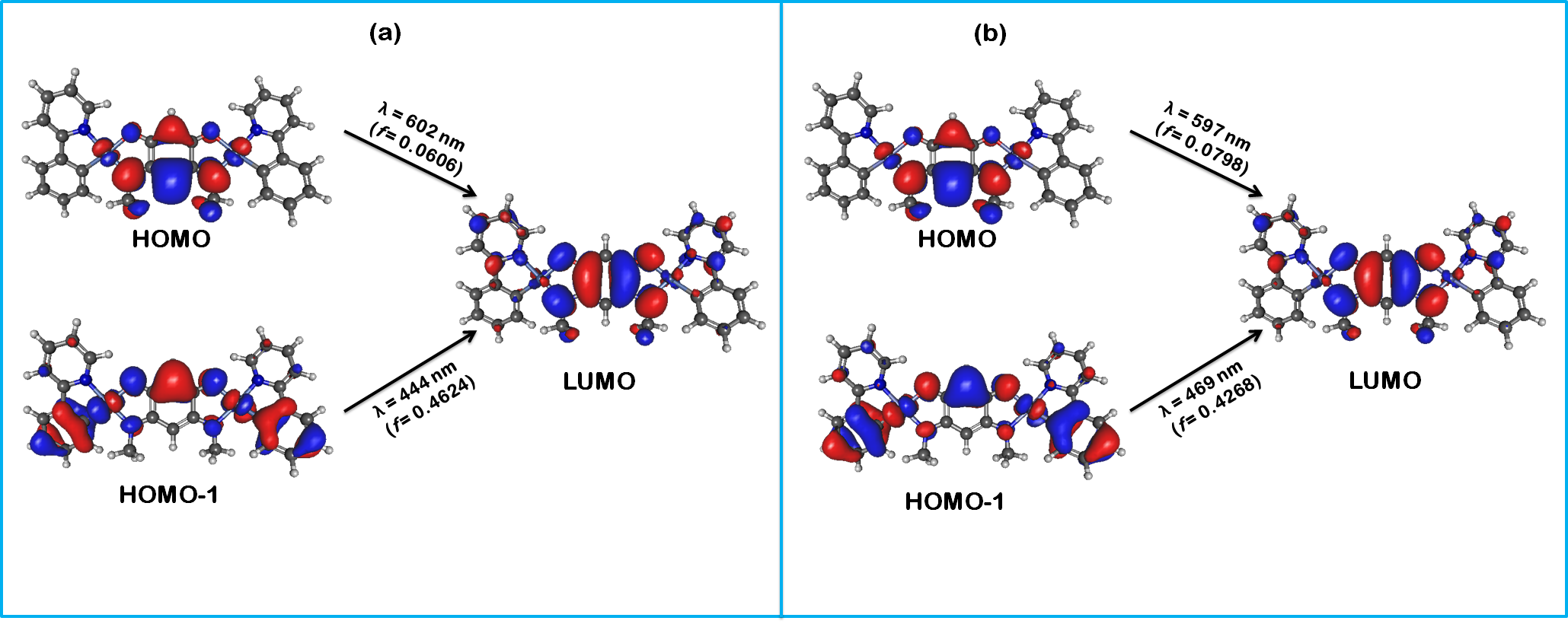}
	\caption{(color online) HOMO, HOMO-1 and LUMO orbital plots for (a)  Ni$^{II}$- and (b) Co$^{II}$-dinuclear molecule obtained from DFT calculations using B3LYP functional. These are the main molecular orbital in which transition take places during the photo-absorption. Transition wave length and corresponding transition probabilities are indicated near arrow marks.}
\end{figure}
Excitation wavelengths and their corresponding oscillator strengths which originates the absorption bands for Ni- and Co-dinuclear molecule are marked with arrows in Fig. 3a. and Fig.3b respectively. Calculated excitation wavelengths and oscillator strengths for the selected transitions of all other molecules is given in SI. 
 
The intense peak in the absorption spectra of  Ni-dinuclear and  Co-dinuclear molecules are  at $\lambda$ = 444 nm  and at $\lambda$ = 469 nm  respectively and are mainly attributed to the (HOMO-1)-LUMO transition in both case. Whereas the HOMO-LUMO transition in both case corresponds to the broad absorption band which are at $\lambda$ = 602 nm  and at $\lambda$ = 596 nm for Ni- and Co-molecule respectively. In case of Ni-molecule another small peak in absorption spectra appears near 440nm wavelength which is due to HOMO-3 to LUMO transition, but the transition probability for such transition is very low(f=0.061).\\
\section{Conclusions}
To conclude, we have studied the electronic structure, magnetic and optical properties  of a set of TM-dinuclear molecules using DFT calculations. The molecules were geometrically optimized in the gas phase and their stability were analyzed by studying vibrational properties.  The proper ground state spin configuration of the tetra-coordinated TM atom in the modeled molecules was predicted by comparing total energies in different possible spin states and observed that the TM atom in Cr-, Mn-, Fe-, Co- and  Ni-dinuclear molecules shows S=2, S=5/2, S=1, S=1/2 and S=0 spin-state, respectively in the ground state. Apart from Mn-dinuclear, the planer structure remain intact in all other molecule and expected to be well adsorb and self assemble on metals substrates would be very suitable for molecular spintronics.  The optical properties of each molecule are calculated using TD-DFT method. We observe for the first time that in case of Cr-, Mn-, Co-, and Ni-dinuclear molecules, the  absorption take place mainly near blue light wavelength range, therefore could be suitable for blue-LED application. Our present work promotes further theoretical and  experimental studies  on the TM-dinculear organometallic molecules for their future application in optoelectronics and molecular spintronics.

\vspace{0.5cm}
\section*{References}
\bibliographystyle{iopart-num}

\end{document}